\documentclass[12pt]{amsart}
\usepackage{amssymb}
\usepackage[mathscr]{eucal}
\usepackage{hyperref}

\allowdisplaybreaks

\def\bref#1{\textbf{\ref{#1}}}

\def\dl#1{\frac{{\d}}{\d #1}}

\def\ddl#1#2{\frac{{\d}}{\d #2}#1}
\def\ddr#1#2{#1\frac{\overleftarrow{\d}}{\d #2}}

\newcommand{\wwedge}{\mathchoice{{\textstyle\bigwedge}}{\bigwedge}{\bigwedge}%
{\bigwedge}}
\def\S{\mathsf{S}}

\def\C{\mathop{\mathsf{C}}\nolimits}
\def\cP{\mathcal P}\def\C{\mathop{\mathsf{C}}\nolimits}

\def\manifold{\mathcal}                               %
\def\manG{\manifold{G}}                               %
\def\manL{\manifold{L}}                               %
\def\manM{\manifold{M}}                               %
\def\manN{\manifold{N}}                               %
\def\manX{\manifold{X}}                               %
\def\ZQ{\manifold{Z}_{\Q}}                            %
\def\Z#1{\manifold{Z}_{#1}}                           %

\def\module{\mathfrak}                                %
\def\mM{\module{M}}                                   %
\def\algebra{\mathfrak}                               %
\def\aA{\algebra{a}}
\def\cA{\algebra{A}}                                  %
\def\QP{QP}                                           %
\def\Q{\boldsymbol{Q}}                                %

\def\structure{\mathscr}                              %
\newcommand{\ideal}[1]{{\structure{I}_{#1}}}          %
\newcommand{\func}[1]{\structure{F}_{#1}}             %
\def\YB{\mathrm{YB}}                                  %
\newcommand{\commut}[2]{\left[#1,\,#2\right]}
\newcommand{\gerst}[2]{\boldsymbol{\{}{}#1{},{}#2{}\boldsymbol{\}}}

\newcommand{\ab}[2]{\left(#1,\,#2\right)}
\newcommand{\pb}[2]{\left\{#1,\,#2\right\}}

\newcommand{\drop}[1]{}
\newcommand{\p}[1]{\mathsf{p}(#1)}

\newcommand{\Hom}[2]{\mathop{\mathsf{Hom}}\nolimits(#1,#2)}

\newcommand{\gh}[1]{{\rm gh}(#1)}

\newcommand{\Lie}[2]{\left[#1\,,\,#2\right]}
\newcommand{\diff}[1]{\frac{\d}{\d #1}}

\renewcommand{\d}{\partial}

\def\tensor{\otimes}
\newcommand{\Vect}[1]{\mathrm{Vect}_{#1}}

\newcommand{\half}{\frac{1}{2}} \newcommand{\thalf}{\tfrac{1}{2}}
\def\bar{\overline}
\def\tilde{\widetilde}

\newcommand{\BFV}{\mathop{\structure{F}}\nolimits}

\swapnumbers

\newtheorem{thm}[subsubsection]{Theorem}

\newtheorem{lemma}[subsubsection]{Lemma}

\newtheorem{prop}[subsubsection]{Proposition}
\newtheorem{Cor}[subsubsection]{Corollary}

\newtheorem{Dfn}[subsection]{Definition}
\newtheorem{dfn}[subsubsection]{Definition}

\theoremstyle{definition}
{\noindent{\hfill\mbox{\rule{.5em}{.5em}}\,}\par\medskip}

\def\NPB{Nucl.\ Phys.\ B}
\def\PLB{Phys.\ Lett.\ B}

\def\CMP{Commun.\ Math.\ Phys.}
\def\IJMPA{Int.\ J.\ Mod.\ Phys.\ A}
\def\PRD{Phys. Rev. D}
\def\JMP{J.\ Math.\ Phys.}

\numberwithin{equation}{section}
\makeatletter 

\@addtoreset{equation}{section}

\def\@secnumfont{\bfseries}
\def\subsubsection{\@startsection{subsubsection}{3}%
  \z@{.5\linespacing\@plus.7\linespacing}{-.5em}%
  {\normalfont\bfseries}}
\def\paragraph{\@startsection{paragraph}{4}%
  \z@\z@{-\fontdimen2\font}%
  \normalfont\bfseries}
\def\subparagraph{\@startsection{subparagraph}{5}%
  \z@\z@{-\fontdimen2\font}%
  \normalfont\bfseries}

\makeatother

\begin{document}
\vfuzz1.2pt
\addtolength{\baselineskip}{4pt}
\addtolength{\parskip}{2pt}
\raggedbottom

\title[BRST Formalism and Zero Locus Reduction]{\hfill{
    \lowercase{\tt hep-th/0001081}}\\[12pt]
  BRST Formalism and Zero Locus Reduction}

\author[Grigoriev]{M.~A.~Grigoriev}
\address{Lebedev Physics Institute, Russian Academy of Sciences}
\author[Semikhatov]{A.~M.~Semikhatov}
\author[Tipunin]{I.~Yu.~Tipunin}

\begin{abstract}
  In the BRST quantization of gauge theories, the zero locus~$\ZQ$ of
  the BRST differential~$\Q$ carries an (anti)bracket whose parity is
  opposite to that of the fundamental bracket.  Observables of the
  BRST theory are in a~$1:1$ correspondence with Casimir functions of
  the bracket on~$\ZQ$.  For any constrained dynamical system with the
  phase space~$\manN_0$ and the constraint surface~$\Sigma$, we prove
  its equivalence to the constrained system on the BFV-extended phase
  space with the constraint surface given by~$\ZQ$.  Reduction to the
  zero locus of the differential gives rise to relations between
  bracket operations and differentials arising in different complexes
  (the Gerstenhaber, Schouten, Berezin--Kirillov, and Sklyanin
  brackets); the equation ensuring the existence of a nilpotent vector
  field on the reduced manifold can be the classical Yang--Baxter
  equation.  We also generalize our constructions to the
  bi-\QP-manifolds which from the BRST theory viewpoint correspond to
  the BRST-anti-BRST-symmetric quantization.
\end{abstract}

\maketitle
\thispagestyle{empty}

\section{Introduction}  
The ``BRST'' quantization of general gauge theories in the Hamiltonian
and Lagrangian formalisms includes the Batalin--Fradkin--Vilkovisky
(BFV)~\cite{[BFV]} and Batalin--Vilkovisky (BV)~\cite{[BV]}
formalisms.  {}From a geometric standpoint, these quantization
formalisms deal with an even or odd
\QP~manifold~$\manN$~\cite{[ASS],[AKSZ]}, i.e., a symplectic or
antisymplectic manifold equipped with a compatible odd vector field
$\Q$ such that~$\Q^2=0$.  This condition is ensured by imposing the
\textit{master equation} on the Hamiltonian function of the vector
field~$\Q$.  In the standard physicists' notation, the respective
equations are
\begin{equation}\label{master0}
  \{\Omega,\Omega\}=0\quad\text{and}\quad
  (S,S)=0,
\end{equation}
where~$\Omega$ (by a widespread abuse of terminology) is the ``BRST
generator'' in the Hamiltonian quantization and~$S$ is the master
action in the Lagrangian quantization.

Under appropriate regularity conditions, the zero
locus~$\ZQ\subset\manN$ of~$\Q=\{\Omega,\cdot\,\}$
(of~$\Q=(S,\cdot\,)$) is an odd Poisson manifold (respectively, a
Poisson manifold)~\cite{[AKSZ],[GST]}, whose geometry captures crucial
information about the theory on~$\manN$.  In this paper, we mainly
concentrate on \textit{even} \QP~manifolds (which correspond to the
BFV quantization and were implicit in~\cite{[BM-Dual]}) because they
have not been considered before; however, we formulate the general
facts about the zero-locus reduction such that they apply to both even
and odd \QP~manifolds.  On an even \QP~manifold,~$\ZQ$ carries an
\textit{anti}bracket; we then show that the equivalence classes of
observables (the cohomology of~$\Q$) are in a~$1:1$ correspondence
with characteristic (Casimir) functions of the antibracket on~$\ZQ$,
and gauge symmetries in the BFV theory on~$\manN$ are Hamiltonian
vector fields on~$\ZQ$.

Moreover, the zero locus~$\ZQ$ of the BFV differential on the extended
phase space is a proper counterpart of the constraint surface in the
following sense.  In geometric terms, a first-class constrained system
can be specified by its phase space (a symplectic manifold~$\manN_0$)
and the constraint surface~$\Sigma$.  On the extended phase space
$\manN$ constructed in the BFV quantization, we can consider the
dynamical system whose constraint surface, by definition, is~$\ZQ$ (in
local coordinates on~$\manN$, the constraints can be chosen as the
components of~$\Q$).  Then the constrained systems~$(\manN_0,\Sigma)$
and~$(\manN,\ZQ)$ \textit{are equivalent}: the respective algebras of
the equivalence classes of observables are naturally isomorphic as
Poisson algebras.

Beyond the BRST context, algebras of functions on \QP~manifolds, which
are differential Poisson algebras (associative supercommutative
algebras endowed with a bracket operation and a differential that is a
derivation of the bracket) can arise from complexes endowed with a
super-commutative associative multiplication and a Gerstenhaber-like
multiplication (``\textit{bracket}''); the differential is then
interpreted as the~$\Q$-structure, and the bracket becomes the
P-structure (the Poisson or the BV bracket on the dual
(super)manifold).  The basic examples are the cohomology complexes of
a Lie algebra~$\aA$ with coefficients in~$\wwedge\aA$ or~$\S\aA$ (the
exterior and symmetric tensor algebras); the general case involves
$L_\infty$ algebras~\cite{[Stasheff]}.

In this algebraic context, reduction to the zero locus can yield
relations between different complexes.  In certain cases, the
zero-locus reduction can be applied \textit{repeatedly}; the equation
ensuring the existence of a nilpotent vector field on the reduced
manifold at the second step of the reduction can be the
classical Yang--Baxter equation (CYBE), in which case the reduction
leads to the well-known Sklyanin and Berezin--Kirillov brackets.

In addition to the usual \QP~manifolds, one can consider
\textit{bi-\QP}~manifolds, which are the geometric counterparts of
bicomplexes, and in physical terms, originate in the BRST--anti-BRST
($Sp(2)$-symmetric/triplectic)
quantization~\cite{[BLT],[BLT-sp2],[BM],[BMS]}.  With two BRST
operators represented by two commuting (odd and nilpotent) vector
fields, bi-\QP~manifolds might be called QQP~manifolds; interestingly
enough, the corresponding zero-locus reduction (to the submanifold on
which both vector fields vanish) results in a ``PP''~manifold, i.e.,
gives rise to a~\textit{bi-Hamiltonian} structure.  A typical example
is obtained by starting with a Lie algebra~$\aA$ and deriving the
second differential from a \textit{co}algebra structure.
Compatibility between two differentials then implies that
$(\aA,\aA^*,\aA \oplus \aA^*)$ is a Manin triple~\cite{[Manin]}.
There also exists an alternative construction of a bi-\QP~manifold
from a \textit{single} Lie algebra structure, which results in
non-Abelian triplectic antibrackets~\cite{[G]} on the space of common
zeroes of the differentials (and thus, the zero locus reduction leads
to a nontrivial relation to the bicomplex used in the extended BRST
symmetry).

\bigskip

This paper is organized as follows. In
Sec.~\bref{subsec:zero-locus-Q}, we recall the main points of the zero
locus reduction on (odd or even) \QP~manifolds.  Symmetries of
\QP~manifolds are reviewed in Sec.~\bref{sec:symmetries}.  In
Sec.~\bref{sec:BFV-symm}, we turn to a more detailed analysis of even
\QP~manifolds corresponding to the BFV quantization.  In
Secs.~\bref{sec:basic}--\bref{sec:fromD2BFV}, we recall several facts
about the BFV formalism in the form that is suitable for what follows.
The results given in~\bref{sec:ZQ}
state the relation between objects in the bulk of the phase space and
on the zero locus submanifold.  We briefly discuss in
Sec.~\bref{sec:gaugesymm} how these results can be restated for the BV
formalism.  In Sec.~\bref{sec:towers}, we consider specific brackets
resulting from the zero-locus reduction.  In Sec.~\bref{bicomplex}, we
study \textit{bi}-\QP~manifolds.

\section{Geometry of \QP~manifolds and zero locus
  reduction}\label{zero-locus} Geometric objects underlying the BRST
quantization are the~$\QP{}$ manifolds.
\begin{Dfn}[\cite{[ASS],[AKSZ]}]\label{def:QP}
  A \QP~manifold is a supermanifold~$\manN$ equipped with a
  bracket~$\gerst{\cdot{}}{\cdot{}}$ such that
  \begin{equation}\label{eq:Jacobi}
    \begin{split}
      \gerst{F}{G}={}&
      -(-1)^{(\p{F}+\kappa)(\p{G}+\kappa)}\gerst{G}{F},\\  
      \gerst{F}{GH}={}&\gerst{F}{G}
      H+(-1)^{(\p{F}+\kappa))\p{G}}G\gerst{F}{H}, \\
      \gerst{F}{\gerst{G}{H}}={}&
      \gerst{\gerst{F}{G}}{H}+\gerst{G}{\gerst{F}{H}}
      (-1)^{(\p{F}+\kappa)(\p{G}+\kappa)},
    \end{split}
  \end{equation}
  for~$F,G,H \in \func{\manN}$ (smooth functions on~$\manN$), and with
  an odd nilpotent vector field~$\Q$, $\Q^2=0$, such that
  \begin{equation}\label{Q-differentiates}
    \Q\gerst{F}{G}
    -\gerst{Q F}{G}-(-1)^{\p{F}+\kappa}\gerst{F}{\Q G}=0,
    \quad F,G\in\func\manN
  \end{equation}
  (where $\p(\cdot)$ is the Grassmann parity).  \QP~manifolds with a
  \emph{Poisson bracket ($\kappa=0$)} are called even, and those with
  an \emph{antibracket ($\kappa=1$)}, odd.
\end{Dfn}
Odd~$\QP{}$ manifolds arise in the BV quantization, and even ones in
the BFV quantization.  Odd \QP~manifolds were introduced
in~\cite{[ASS]} and were studied in~\cite{[AKSZ],[GST]}.  In
most of our definitions, \QP~manifolds can be either even or odd; in
Sec.~\bref{sec:BFV-symm}, however, we concentrate on even
\QP~manifolds, which have not been given enough attention previously.

\subsection{The zero locus of $\Q$}\label{subsec:zero-locus-Q} In what
follows,~$\ZQ$ denotes the zero locus of the odd vector field~$\Q$ on
a~$\QP{}$ manifold~$\manN$.  We assume~$\ZQ$ to be a nonempty smooth
submanifold and denote by~$\ideal{\ZQ}\subset \func\manN$ \textit{the
  ideal of smooth functions vanishing on~$\ZQ$}.

The odd vector field~$\Q$ is called \textit{regular} if each function
$f\in\ideal{\ZQ}$ can be represented as
\begin{equation}\label{eq:regular}
  f=\sum_\alpha f_\alpha\,\Q \Gamma^\alpha,
\end{equation}
with some~$f_\alpha,\,\Gamma^\alpha\in\func\manN$ (i.e., if the
components of~$\Q$ generate~$\ideal{\ZQ}$).  We say that a submanifold
$\manL\subset\manN$ is \textit{coisotropic} if
\begin{equation}
  \gerst{\ideal{\manL}}{\ideal{\manL}}\subset\ideal{\manL}.
\end{equation}
\begin{lemma}\label{fact:coisotropic}
  If~$\Q$ is regular, $\ZQ$ is a coisotropic submanifold of the \QP{}
  manifold~$\manN$.
\end{lemma}

\begin{proof}
  Let $f,g \in \func\manN$ vanish on~$\ZQ$.  Using
  representation~\eqref{eq:regular}, the Leibnitz rule, 
  Eq.~\eqref{Q-differentiates}, and nilpotency
  of~$\Q$, we see that~$\gerst{f_\alpha
    (\Q\Gamma^\alpha)}{(\Q\Gamma^\beta) g_\beta}|_{\ZQ}=
  (f_\alpha\gerst{\Q\Gamma^\alpha}{\Q \Gamma^\beta} g_\beta
  )|_{\ZQ}=0$.
\end{proof}
\noindent
In what follows, we assume~$\ZQ$ to be coisotropic even in those
cases where~$\Q$ is not regular.

The algebra~$\func{\ZQ}$ of smooth functions on~$\ZQ$ is the quotient
${\func\manN}/{\ideal{\ZQ}}$.  We then have
\begin{lemma}\label{lemma:binary}
  There is a well-defined binary operation given by
  ${\gerst{\;}{\;}}_{\Q}:{}\func{\Z\Q} \times \func{\Z\Q} \to
  \func{\Z\Q}$
  \begin{equation}\label{binary}
    {\gerst{f}{g}}_{\Q}=\gerst{F}{\Q\,G}|_{\Z\Q},\qquad
    f,g \in\func{\Z\Q},\quad
    F,G \in \func\manN,\quad F|_{\Z\Q}=f,\quad G|_{\Z\Q}=g,
  \end{equation}
  where~$F$ and~$G\in\func\manN$ are viewed as representatives of
  functions on~$\ZQ$. It makes~$\ZQ$ into a Poisson manifold.
\end{lemma}
\noindent
The proof is a straightforward generalization of a proof given
in~\cite{[GST]}.  It is obvious that the parity of the induced bracket
on~$\ZQ$ is opposite to the parity of the~$\gerst{\,}{\,}$ bracket
on~$\manN$.  An important characteristic of the differential~$\Q$ is
the homology of the linear operators~$\Q_p:T_p\manN \to T_p\manN$,
$p\in\ZQ$, defined as follows.  We consider the tangent space
$T_p\manN$ as the quotient of the vector fields~$\Vect\manN$ modulo
those that vanish at~$p$. Then
\begin{equation}
  \Q_p (x)=(\commut{\Q}{X})|_p,\quad
  X\in\Vect\manN,\quad x=X_p\in T_p\manN.
\end{equation}
This operation is well-defined once~$\Q$ vanishes at~$p$.
\begin{dfn}\label{def:proper}
  A \QP~manifold~$\manN$ is called \emph{proper} if the homology of
  the linear operator~$\Q_p:T_p\manN\to T_p\manN$ is trivial at each
  point~$p\in\ZQ$.
\end{dfn}
\noindent
This definition is equivalent to the one given in~\cite{[AKSZ]}
(and~\cite{[GST]}), but uses only invariant notions (in local
coordinates~$\Gamma^A$, we would have~$(\Q_p\,x)^A=(-1)^{\p{x}+1}
x^B\frac{\d \Q^A}{\d \Gamma^B}$.).  We now have
\begin{prop}[\cite{[AKSZ],[GST]}]
  Let~$\manN$ be a proper \QP~manifold with a nondegenerate bracket.
  Then $\ZQ$ is (anti)sym\-plectic with respect to the induced
  bracket~\eqref{binary}.
\end{prop}

One can replace~$\ZQ$ with a submanifold that still is coisotropic.
As a straightforward generalization of~\bref{lemma:binary}, we have
\begin{thm}\label{thm:general-zero}
  Let~$\manN$ be a \QP~manifold and~$\manL\subset\ZQ\subset\manN$ a
  coisotropic submanifold of~$\manN$.  Then~$\manL$ is a Poisson
  manifold\footnote{By Poisson manifolds, we mean those with either an
    even bracket or an antibracket.}  with the Poisson structure given
  by
  \begin{equation}\label{eq:LPB}
    {\gerst{f}{g}}_{\Q}={\gerst{F}{\Q\,G}}|_\manL,\qquad
    f,g \in\func\manL,\quad
    F,G \in \func\manN,\quad  F|_\manL=f,\quad G|_\manL=g.
  \end{equation}
\end{thm}
\begin{proof}
  It is easy to see that~\eqref{eq:LPB} does not depend on the choice
  of representatives~$F,G\in \func\manN$ of~$f,g \in\func\manL$.  The
  Jacobi identity and the Leibnitz rule follow in the same way as for
  the bracket in Eq.~\eqref{binary}, see~\cite{[GST]}.
\end{proof}

\subsection{Symmetries of  \QP~manifolds~\cite{[GST]}}
\label{sec:symmetries}
We now recall several basic facts about symmetries of \QP~structures
on a manifold.
\begin{dfn}\label{dfn:QP-symmetry}
  A vector field~$X$ on a \QP~manifold{}~$\manN$ is called a symmetry
  of~$\manN$ if it commutes with~$\Q$ and is a Poisson vector field,
  i.e.,
  \begin{equation}\label{eq:symmetry}
    X\gerst{F}{G} - \gerst{X F}{G} -
    (-1)^{(\p{F}+\kappa)\p{X}}\gerst{F}{X G}=0,\quad F,G\in
    \func\manN.
  \end{equation}
  Symmetries of the form~$X=\gerst{\Q F}{\;\cdot\;}$ (with
  $F\in\func{\manN}$) are called \emph{trivial}.
\end{dfn}

The Lie algebras of symmetries and trivial symmetries behave in a very
regular manner under the restriction to~$\ZQ$.
\begin{prop}\label{fact:restriction}
  Let~$X$ be a symmetry of~$\manN$.  Then~$X$ restricts to~$\ZQ$ and
  its restriction~$x$ is a Poisson vector field on~$\ZQ$ with respect
  to the bracket \eqref{eq:LPB} on~$\ZQ$, namely
  \begin{equation}\label{eq:ZLB-preserving}
    x\gerst{F}{G}_{\Q}- \gerst{x F}{G}_{\Q}-
    (-1)^{(\p{F}+\kappa+1)\p{X}} \gerst{F}{x G}_{\Q}=0,\quad
    F,G\in\func{\ZQ}.
  \end{equation}
  If in addition~$X=\gerst{\Q H}{\cdot\,}$ is a trivial symmetry,
  $x$ is a Hamiltonian vector field with respect to the
  $\gerst{~}{~}_{\Q}$ bracket.  
\end{prop}
\begin{proof}
  Any symmetry~$X$ restricts to~$\ZQ$ because~$X F|_{\ZQ}=0$ for any
  $F$ vanishing on~$\ZQ$. Indeed, every such function can be
  represented as~$F=F_\alpha\cdot\Q \Gamma^\alpha$ with some functions
  $F_\alpha$ and~$\Gamma^\alpha$, provided $\Q$ is regular.  Because
  $[X,\Q]\!=\!0$, we have $X F|_{\ZQ}=\!((X F_\alpha)(\Q
  \Gamma^\alpha))|_{\ZQ} + (-1)^{\p{X}(\p{F_\alpha}+1)F_\alpha}(\Q X
  \Gamma^\alpha)|_{\ZQ}=0$.  Equation~\eqref{eq:ZLB-preserving}
  immediately follows from the definition of the zero locus bracket
  and the definition of symmetries.  If in addition~$X=\gerst{\Q H
    }{\cdot}$ is a trivial symmetry, for any function~$f \in \func\ZQ$
  we have
 \begin{equation}
    xf=X|_{\ZQ}f=\gerst{\Q H}{F}|_{\ZQ}=(-1)^{p(H)+\kappa+1}
    \,\gerst{H|_{\ZQ}}{f}_{\Q},
 \end{equation}
 where~$F \subset \func\manN$ is a lift of~$f$ (i.e.,~$f=F|_{\ZQ}$)
 and~$\kappa$ is the parity of the~$\gerst{\,}{\,}$ bracket.  Thus,
 $x=X|_{\ZQ}$ is a Hamiltonian vector field with respect to the
 bracket~$\gerst{\,}{\,}_{\Q}$.
\end{proof}

\section{Observables, gauge symmetries, and zero locus reduction
  in BFV and BV quantizations} \label{sec:BFV-symm} We now consider
the embedding of a constrained system into the BFV extended theory
with the BRST charge~$\Omega$ and study the ``on-shell'' gauge
symmetries in the two descriptions of the same theory.  In the Dirac
(``non-extended'') formalism, the on-shell gauge symmetries are those
nonvanishing on the constraint surface, and in the BFV extended
formalism, these are symmetries nonvanishing on the zero locus~$\ZQ$.
We show that the former are mapped into the latter such that the
equivalence classes of observables in the original theory are mapped
into equivalence classes of observables in the BFV theory (the latter
can be considered as gauge invariant functions on~$\ZQ$).  In this
sense, \textit{the zero locus~$\ZQ$ plays the role of a constraint
  surface in the BFV theory}.  We concentrate on the BFV case, where
we assume the phase space to be finite-dimensional; reformulation of
our results for the BV quantization, although straightforward at the
formal level, requires some care because the BV configuration space of
any realistic model is infinite-dimensional
(see~\bref{sec:gaugesymm}).

\subsection{A reminder on constrained dynamics}\label{sec:basic} We
begin with recalling several basic facts about constrained dynamics in
the form that will be suitable in what follows.

\subsubsection{Basics of the Dirac constrained
  dynamics}\label{sec:initial} We consider a first-class constrained
Hamiltonian system, defined on a phase space (symplectic
manifold)~$\manN_0$ with the \textit{constraints} $T_\alpha$
(functions on~$\manN_0$) such that
\begin{equation}\label{eq:Lie}
  \pb{T_\alpha}{T_\beta}=U^\gamma_{\alpha\beta} T_\gamma,
\end{equation}
where $\pb{\;}{\,}$ is the Poisson bracket on~$\manN_0$.  For
simplicity, we assume the first-class constraints~$T_\alpha$ to be
irreducible.  Let~$\Sigma$ denote the \textit{constraint
  surface}~$T_\alpha=0$.  A geometrically invariant way to specify a
first-class constrained system is to fix the pair $(\manN_0,\Sigma)$
(a symplectic manifold and a coisotropic submanifold).  Different
choices for~$T_\alpha$ then give different generators of the ideal of
functions vanishing on~$\Sigma$.

By definition, an \textit{observable} is a function on~$\manN_0$
satisfying~$\gerst{A}{T_\alpha}|_\Sigma=0$.  Under the standard
regularity conditions, each function vanishing on~$\Sigma$ is
proportional to the constraints, and therefore,
\begin{equation}\label{eq:cobserv}
  \pb{A}{T_\alpha}=A^\beta_\alpha T_\beta
\end{equation}
for some functions~$A^\alpha_\beta$.  Observables vanishing on
$\Sigma$ are called \textit{trivial}.  Two observables are called
\textit{equivalent} if they differ by a trivial observable.  The space
of equivalence classes of observables is a Poisson algebra, i.e., is
closed under multiplication and under the Poisson bracket (these
operations are well-defined on the equivalence classes via
representatives).  This algebra can be conveniently thought of as a
subalgebra in the algebra of functions on~$\Sigma$.

\textit{Infinitesimal gauge transformations}, or \textit{gauge
  symmetries}, are the Hamiltonian vector fields
$X_0=\pb{\phi_0}{\cdot\,}$, where~$\phi_0=\phi^\alpha_0 T_\alpha$ is a
trivial observable.  Gauge symmetries form a Lie algebra with respect
to the commutator.  For any observable~$A$ and a gauge symmetry
$X_0=\pb{\phi_0}{\cdot\,}$, we have
\begin{equation}
  X_0 A=\pb{\phi_0}{A}=\pb{\phi^\alpha_0 T_\alpha}{A}=
  \phi_0^{\alpha}\pb{T_\alpha}{A}+T_\alpha \pb{\phi_0^\alpha}{A},
\end{equation}
which vanishes on~$\Sigma$ because~$A$ is an observable.  Therefore,
gauge symmetries preserve equivalence classes of observables.

By the \textit{on-shell gauge symmetries}, we mean the equivalence
classes of gauge symmetries modulo those vanishing on the constraint
surface~$\Sigma$.  On-shell gauge symmetries can also be viewed as a
subalgebra in the algebra of vector fields on~$\Sigma$.  Equivalence
classes of observables (viewed as functions on~$\Sigma$) are then
represented by functions annihilated by on-shell gauge symmetries.

\subsubsection{Basics of the BFV/BRST approach}
\label{sec:BFV-observables} In the BFV quantization, the extended
phase space~$\manN$ is an even \QP~manifold whose $\Q$-structure is
given by~$\Q=\pb{\Omega}{\cdot\,}$, where~$\Omega$ is a function
on~$\manN$ (called the BRST charge)
satisfying~$\pb{\Omega}{\Omega}=0$.  In applications, the BFV extended
phase space is usually equipped with an additional structure, the
ghost charge~$G\in\func{\manN}$.  Functions with a definite ghost
number are eigenfunctions of the ghost number operator
\begin{equation}\label{eq:BFV-ghost-number}
  g=\pb{G}{\cdot\,},
\end{equation}
corresponding to integer eigenvalues.  The BRST charge is required to
have the ghost number~$1$,
\begin{equation}\label{eq:Omega-ghost-number}
  \pb{G}{\Omega}=\Omega.
\end{equation}

We now consider a \QP~manifold~$\manN$ that is not necessarily
constructed via the BFV prescription; however, we refer to the objects
on~$\manN$ as BFV ones because the applications in what follows will
be to the case where~$\manN$ does result from the BFV construction.
This also helps to distinguish between observables and symmetries on
the \QP~manifold and those in the initial
theory~(Sec.~\bref{sec:initial}), with `BFV' used to refer to the
former.

\textit{A BFV observable}~$A$ is a function on the \QP~manifold
$\manN$ satisfying
\begin{equation}\label{eq:BFV-observ}
  \Q A=\pb{\Omega}{A}=0,\qquad \gh{A}=0.
\end{equation}
The~$\Q$-exact BFV observables are called \textit{trivial}.  Two BFV
observables~$A$ and~$\tilde A$ are equivalent if~$A-\tilde A=\Q B$ for
some function~$B$; the equivalence classes of observables are then the
cohomology of~$\Q$ in the ghost number zero.  The algebra of BFV
observables is a Poisson algebra (multiplication and the Poisson
bracket can be defined via representatives).

A vector field~$X$ is called a \textit{BFV gauge symmetry} if
$X=\pb{\Q H }{\cdot}$ for some function~$H$ with~$\gh{X}=\gh{\Q H}=0$
(these are trivial symmetries (see~\bref{dfn:QP-symmetry}) of the
corresponding \QP~manifold).  In other words, BFV gauge symmetries are
the Hamiltonian vector fields generated by trivial BFV observables.
If~$A$ is an observable and~$X=\pb{\Q H}{\cdot}$ a BFV gauge symmetry,
we see that~$X A=\pb{\Q H}{A}=\Q\pb{H}{A}$ is a trivial observable,
i.e., BFV gauge symmetries preserve the equivalence classes of BFV
observables.

\subsection{From Dirac to the BFV formulation of a constrained
  system}\label{sec:fromD2BFV} Formal similarities between the Dirac
and BFV formalisms are summarized in Table~\ref{tab:summary}.
\begin{table}[bt]
  \begin{center}\renewcommand{\arraystretch}{1.4}
    \begin{tabular}{|l|c|c|}
      \hline
      & Dirac (Sec.~\bref{sec:initial})& BFV
      (Sec.~\bref{sec:BFV-observables})\strut\\
      \hline
      observables& $A_0$, $\pb{A_0}{T_\alpha}=
      A^\beta_\alpha T_\beta$ &
      $\Q A=\pb{\Omega}{A}=0$, $\gh{A}=0$\\
      trivial observables &$(A_0)|_\Sigma=0$ & $A=\Q B$\\
      equivalent observables &$A_0\sim A_0 + a^\alpha T_\alpha$&
      $A\sim A + \Q B$\\
      gauge symmetries &$X_0=\pb{\phi^\alpha_0 T_\alpha}{\cdot\,}$&
      $X=\pb{\Q H }{\cdot}$, $\gh{X}=0$\\
      \hline
    \end{tabular}
  \end{center}\bigskip\caption{\label{tab:summary}}
\end{table}
We now make contact between~\bref{sec:initial}
and~\bref{sec:BFV-observables} by taking the extended phase space
$\manN$ and the BRST charge~$\Omega$ to be those arising in the BFV
formalism from a given first-class constrained
system~$(\manN_0,\Sigma)$.  As before, the constraints~$T_\alpha \in
\func{\manN_0}$ are taken to be irreducible; to construct the BFV
formalism, one then introduces ghosts~$c^\alpha$,
with~$\gh{c^\alpha}=1$, $\p{c^\alpha}=\p{T_\alpha}+1$ and their
conjugate momenta $\cP_\alpha$,
\begin{equation}\label{eq:ghost-PB}
  \pb{c^\alpha}{\cP_\beta}=\delta^\alpha_\beta,
\end{equation}
with~$\gh{\cP_\alpha}=-1$, $\p{\cP_\alpha}=\p{T_\alpha}+1$.  The
extended phase space~$\manN$ is the direct product of~$\manN_0$ with
the superspace spanned by $c^\alpha$ and~$\cP_\alpha$.\footnote{When
  the constraints are defined \textit{locally}, the extended phase
  space is a vector bundle over the original phase space, as, for
  example, in~\cite{[GL]}.}  The Poisson bracket on~$\manN$ is the
product Poisson bracket of that on~$\manN_0$ and~\eqref{eq:ghost-PB}.

One introduces the ghost charge (where we assume the constraints to be
bosonic to avoid extra sign factors)
\begin{equation}
  G=c^\alpha\cP_\alpha,\qquad \pb{G}{c^\alpha}=c^\alpha,\quad
  \pb{G}{\cP_\alpha}=-\cP_\alpha.
\end{equation}
The BRST charge~$\Omega$ is an odd function defined by the condition
that it has the ghost number~1 and satisfies
\begin{equation}\label{eq:BFV-master}
  \pb{\Omega}{\Omega}=0
\end{equation}
with the boundary condition
\begin{equation}\label{more}
  \Omega=c^\alpha T_\alpha+\ldots,
\end{equation}
where~$~\ldots~$ means higher-order terms in the ghost momenta.  It is
well known~\cite{[BFV],[H-omega],[BLT],[Henn]} that under standard
assumptions, the BRST charge~$\Omega$ exists for any constrained
system.  Up to the first order in~$\cP_\alpha$, one has
\begin{equation}\label{more2}
  \Omega=c^\alpha T_\alpha-\thalf \cP_\gamma
  U^\gamma_{\alpha\beta}c^\alpha c^\beta + \text{\ldots},
\end{equation}
where the structure functions are those from~\eqref{eq:Lie}.

As regards observables, the following statement is well
known~\cite{[BFV],[H-omega]} (see also~\cite{[Henn]}).
\begin{prop}\label{fact:observables-embedding}
  The algebra of the equivalence classes of observables on~$\manN_0$
  and the algebra of the equivalence classes of BFV observables (the
  cohomology of~$\Q$ in the ghost number zero) on the extended phase
  space~$\manN$ are 
  isomorphic as Poisson algebras.
\end{prop}
This means that if $A_0 \in \func{\manN_0}$ is an observable of the
constrained system on~$\manN_0$, there exists a BFV observable~$A \in
\func\manN$ with~$\gh{A}=0$ such that
\begin{equation}
  A|_{\manN_0}=A_0.
\end{equation}
Moreover, two BFV observables corresponding to the same
observable~$A_0$ differ by a trivial BFV observable.  If in
addition~$A_0$ is a trivial observable, it follows
that~$A=\pb{\Omega}{B}$.  The Poisson bracket on~$\manN$ induces a
bracket on the cohomology of~$\Q$, and one has
\begin{equation}
  \{A,B\}\bigr|_{\manN_0} = \{A_0,B_0\}.
\end{equation}
The isomorphism between the BRST cohomology in the ghost number zero
and the algebra of equivalence classes of observables of the
constrained system on~$\manN_0$ is given by the restriction of
representatives to the initial constrained surface
$\Sigma\subset\manN_0\subset\manN$ (recall that equivalence classes of
observables are gauge invariant functions on $\Sigma$).

It also follows from~\bref{fact:observables-embedding} that because
gauge symmetries of the initial system $(\manN_0,\Sigma)$ are
generated by trivial observables, each gauge symmetry can be lifted to
a BFV gauge symmetry.

\subsection{Zero locus $\ZQ$ in the BFV theory, the general case}
\label{subsec:general-QP} We now consider an even \QP-manifold 
$\manN$ that is not necessarily constructed by the BFV procedure for a
constrained system.  We assume $\manN$ to be symplectic, and the odd
nilpotent vector field $\Q$ to be regular in the sense
of~\bref{eq:regular}.  The zero locus~$\ZQ$ is thus a coisotropic
submanifold of $\manN$.  Because each trivial BFV observable~$A=\Q B$
vanishes on~$\ZQ$, each cohomology class uniquely determines a
function on~$\ZQ$.  Thus, there is a mapping
\begin{equation}\label{H2ZQ-mapping}
  H^0_{\Q}\to\func{\ZQ}
\end{equation}
from the space of inequivalent observables to functions on~$\ZQ$.  

In what follows, we say that a statement holds locally if it is true
in every sufficiently small neighbourhood.
Mapping~\eqref{H2ZQ-mapping} is locally an embedding in view of the
following proposition.
\begin{prop}\label{fact:vanishing-trivial}
  Let~$\Q=\pb{\Omega}{\cdot}$ be regular in the sense
  of~\bref{subsec:zero-locus-Q}.  Locally, each BFV observable
  vanishing on~$\ZQ$ is 
  a trivial BFV observable.
\end{prop}
\begin{proof}
  Let~$A$ be an observable vanishing on~$\ZQ$, i.e.,~$\Q
  A=0$,~$A|_{\ZQ}=0$.  We must show that~$A=\Q X$ in a sufficiently
  small neighbourhood~$U$ of any point~$p\in \ZQ$. It is well known
  that locally there exists a coordinate system
  $p_i,q^j,p_\alpha,q^\beta,c^\alpha,\cP_\beta$ on~$\manN$ such that
  \begin{equation}\label{eq:abelian-Omega}
    \begin{split}
      &\Omega=p_ic^i,\\
      \pb{q^i}{p_j}=\delta^i_j,\quad &
      \pb{q^\alpha}{p_\beta}=\delta^\alpha_\beta,\quad
      \pb{c^\alpha}{\cP_\beta}=\delta^\alpha_\beta.\quad
    \end{split}
  \end{equation}
  Since the function~$A$ vanishes on~$\ZQ$, it can be represented as
  \begin{equation}
    A=A^\alpha p_\alpha + A_\alpha c^\alpha.
  \end{equation}
  Now the odd vector field~$\Q$ becomes
  \begin{equation}
    \Q=-c^\alpha\dl{q^\alpha}+p_\alpha\dl{\cP_\alpha},
  \end{equation}
  and can be considered as the exterior differential under the
  identification~$c^\alpha=-dq^\alpha$, $p_\alpha=d\cP_\alpha$, while
  $A$ becomes a 1-form.  The assertion immediately follows from the
  super analogue of the Poincar\'e lemma in~$U$.
\end{proof}

\subsubsection{Embedding $H^0_{\Q}$ into functions on $\ZQ$.} 
As before, $\ZQ$ is the zero locus submanifold of~$\Q=\pb{\Omega}{\,\cdot}$
We recall from~\bref{fact:restriction} that each BFV gauge
symmetry~$X$ can be restricted to~$\ZQ$ and the restriction
$x=X|_{\ZQ}$ is a~${\pb{~}{~}}_{\Q}$-Hamiltonian vector field on
$\ZQ$.  The image of BFV gauge symmetries under the restriction
to~$\ZQ$ is called the algebra of the \textit{on-shell BFV
  symmetries}.  The functions on~$\ZQ$ that are annihilated by the
on-shell BFV symmetries are then the characteristic functions of the
${\pb{~}{~}}_{\Q}$ antibracket on $\ZQ$.\footnote{We recall that a
  function $f$ on the (odd) Poisson manifold $\manN$ is said to be a
  characteristic (Casimir) function of an (odd) Poisson bracket
  $\pb{\,}{}$ if $\pb{f}{H}=0$ for any function $H$.}

Because BFV observables are annihilated by BFV gauge symmetries, the
restriction to~$\ZQ$ maps BFV observables into characteristic
functions of~${\pb{~}{~}}_{\Q}$.  For the equivalence classes of BFV
observables (the cohomology of~$\Q$), this mapping is certainly an
embedding locally.  It is also an isomorphism in the important case of
a BFV \QP~manifold considered in~\bref{sec:ZQ}.  Locally, we choose
flat coordinates in some neighborhood $U$ of a point $p\in\ZQ$ and use
explicit form \eqref{eq:abelian-Omega} of the BRST charge $\Omega$ and
the Poisson bracket to arrive at
\begin{thm}\label{thm:1:1-general}
  Locally, the equivalence classes of BFV observables (the cohomology
  of~$\Q$) are in a~$1:1$ correspondence with characteristic functions
  of the 
  ${\pb{~}{}}_{\Q}$ antibracket on~$\ZQ$.
\end{thm}

We note that in one direction, this statement holds in general (i.e.,
not only locally) because for any observable $A$, we have
\begin{equation}
  \pb{f}{A|_{\ZQ}}_{\Q}=\pb{F}{\Q A}|_{\ZQ}=0,
\end{equation}
where~$F \in \func\manN$ is the lift of a function~$f\in\func{\ZQ}$
and~$A|_{\ZQ}$ is the image of~$A$ under~\eqref{H2ZQ-mapping}.
Thus,~$A|_{\ZQ}$ is a characteristic function of the antibracket
$\pb{~}{\,}_{\Q}$ on~$\ZQ$.

The ``$\ZQ$-based'' view on the BFV formalism developed here can be
expressed as follows.  \textit{Any even~$\QP$~manifold~$\manN$ gives
  rise to the constrained system $(\manN,\ZQ)$, i.e., a constrained
  system whose phase space is~$\manN$ and the constrained surface
  is~$\ZQ$}.  We recall from~\bref{sec:initial} that gauge
transformations and the algebra of observables can be reconstructed if
a first-class constrained system is specified in geometric terms, via
its phase space (a symplectic manifold) and the constraint surface (a
coisotropic submanifold).  We now take this pair to be $(\manN, \ZQ)$
(with $\ZQ$ being coisotropic in view of~\bref{fact:coisotropic}).  In
local coordinates, the constraints are the components of~$\Q$; in a
neighborhood $U\subset \manN$, the following statement is obvious in
the special coordinates in which $\Omega$ and $\pb{}{}_{\Q}$ are given
by~\eqref{eq:abelian-Omega}.
\begin{thm}\label{thm:equivalence-general}
  On a \QP~manifold $\manN$, the constrained system $(\manN,\ZQ)$ is
  locally equivalent to the BFV theory on the extended phase space
  $\manN$ with the BRST charge $\Omega$ (i.e., the respective algebras
  of equivalence classes of observables are isomorphic as Poisson
  algebras).
\end{thm}

\noindent
In a more physical language, the equivalence can be reformulated by
saying that the two constrained dynamics are equivalent.

The above considerations show that BFV observables are related
to~$\ZQ$ in the same way as observables in the initial theory
(Sec.~\bref{sec:initial}) are related to the constraint
surface~$\Sigma$.  This allows us to interpret~$\ZQ$ as the extended
constraint surface. In the general case, this
correspondence takes place at the local level
only.

\subsection{Zero locus $\ZQ$ in the BFV formulation of a constrained
  system} \label{sec:ZQ} We now concentrate on the important case
where the \QP~manifold under consideration is a BFV extended phase
space obtained by the BFV procedure from a given constrained system
$(\manN_0,\Sigma)$.
\begin{prop}\label{fact:Sigma-embedding}
  The initial constraint surface~$\Sigma\subset \manN_0$ is a
  submanifold of the zero locus~$\ZQ \subset \manN$ of the BRST
  differential~$\Q=\pb{\Omega}{\cdot\,}$.
\end{prop}
\begin{proof}
  We restrict ourselves to an irreducible theory with constraints
  $T_\alpha$ (although the statement also is true for reducible
  constraints); the structure of the BRST charge is then given
  by~\eqref{more}.  Considered as a submanifold in $\manN$, the
  initial phase space $\manN_0$ is determined by the
  equations~$c^\alpha=0$ and $\cP_\alpha=0$.  It follows
  from~\eqref{more} and from $\gh{\Omega}=1$ that the zero locus~$\ZQ$
  is determined by the equations
  \begin{equation}
    T_\alpha+\ldots=0,\qquad \ldots=0,
  \end{equation} 
  where~$\ldots$ denotes terms vanishing on~$\manN_0$.  Then the
  intersection~$\ZQ \cap \manN_0$ (considered as a submanifold in
  $\manN_0$) is determined by the equations~$T_\alpha=0$, and
  therefore, coincides with the initial constraint surface~$\Sigma$.
  Thus, $\Sigma$ is a submanifold in~$\ZQ$.
\end{proof}

The zero locus can be described somewhat more explicitly if we recall
that in the BFV formalism, functions on the extended phase space are
formal power series in the ghost variables~$c^\alpha$ and
$\cP_\alpha$.  This means that $\ZQ$ is actually determined by the
equations
\begin{equation}\label{eq:ZQ-equations}
  T_\alpha=0, \qquad c^\alpha=0.
\end{equation}
This, in its turn, gives an explicit construction of the
antibracket~$\pb{\;}{}_{\Q}$ on~$\ZQ$.  Let~$y^i$ be local coordinates
on~$\Sigma$. Then~$y^i$ and~$\cP_\alpha$ can be considered as local
coordinates on~$\ZQ$.  Evaluating~\eqref{binary}, we now obtain
\begin{equation}\label{eq:explicit-AB}
  \pb{y^i}{y^j}_{\Q}=0,\quad
  \pb{\cP_\alpha}{y^i}_{\Q}=R^i_\alpha(y),\quad
  \pb{\cP_\alpha}{\cP_\beta}_{\Q}=
  U^\gamma_{\alpha\beta}(y)\cP_\gamma,
\end{equation}
where~$R^i_\alpha(y)=\pb{T_\alpha}{y^i}|_\Sigma$ and
$U^\gamma_{\alpha\beta}(y)=U^\gamma_{\alpha\beta}|_\Sigma$ with
$U^\gamma_{\alpha\beta}$ from~\eqref{more2}.

Using the explicit form~\eqref{eq:explicit-AB} of the antibracket on
$\ZQ$, it is easy to describe its characteristic functions in terms of
the initial constraint surface $\Sigma$.  The following statement is
obvious for irreducible constraints $T_\alpha$ and can be easily
generalized to reducible constraints.
\begin{prop}\label{fact:explicit-description}
  Characteristic functions of the antibracket $\pb{\,}{}_{\Q}$ are in
  a $1:1$ correspondence with gauge invariant functions on $\Sigma$.
\end{prop}

On a \QP~manifold constructed in accordance with the BFV
  prescription, the relation between the BRST cohomology and the
  geometry of the extended constrained surface $\ZQ$ can be made more
  precise than in the previous section.  In particular, the respective
  counterparts of statements~\bref{fact:vanishing-trivial},
  \bref{thm:1:1-general}, and \bref{thm:equivalence-general}
  \textit{hold globally}. We first see that~\eqref{H2ZQ-mapping} is an
  embedding.
\begin{prop}\label{fact:BFV-vanishing-trivial}
  On a \QP~manifold $\manN$ constructed in the BFV formalism,
    each BFV observable that vanishes on~$\ZQ \subset \manN$ is a
    trivial BFV observable.
\end{prop}
\begin{proof}
  Let $A$ be a BFV observable and~$A|_{\ZQ}=0$. According
  to~\bref{fact:Sigma-embedding}, $\Sigma\subset\ZQ$.  Then
  $A|_{\ZQ}=0$ implies~$A|_\Sigma=0$ (a trivial observable).
  By~\bref{fact:observables-embedding}, $A$ is a trivial BFV
  observable.
\end{proof}

We now consider the \QP~manifold constructed in the BFV formalism.
Combining~\bref{fact:BFV-vanishing-trivial} with the argument given
after~\bref{thm:1:1-general} proves the next theorem in one direction;
the other direction follows because each characteristic function on
$\ZQ$ can be lifted to a BFV observable on~$\manN$,
see~\bref{fact:observables-embedding}
and~\bref{fact:explicit-description}.
\begin{thm}\label{thm:1:1}
  Equivalence classes of BFV observables (the cohomology of~$\Q$ with
  the ghost number zero) on the BFV \QP~manifold are in a~$1:1$
  correspondence with characteristic functions of the zero locus
  antibracket on~$\ZQ$.
\end{thm}

As in~\bref{thm:1:1-general}, we now consider the extended phase space
of the BFV formulation as the phase space of a ``new'' constrained
system determined by the constraint surface~$\ZQ$.  With~$\manN$ in
its turn obtained from a constrained dynamical
system~$(\manN_0,\Sigma)$ in accordance with the BFV formalism, we
have a \textit{global} version of~\bref{thm:equivalence-general}.
\begin{thm}\label{thm:1:1-global}
  Let $\manN$ be a \QP~manifold constructed in the BFV formalism.  The
  constrained system determined by the pair $(\manN, \ZQ)$ is
  equivalent to the BFV theory on~$\manN$ (i.e., the respective
  algebras of equivalence classes of observables are isomorphic as
  Poisson algebras).
\end{thm}
Combining this with~\bref{fact:observables-embedding}, we obtain a
remarkable relation between the constrained systems specified by the
respective pairs~$(\manN_0,\Sigma)$ and $(\manN,\ZQ)$:
\begin{Cor}
  The constrained systems~$(\manN_0,\Sigma)$ and~$(\manN,\ZQ)$ are
  equivalent (the respective algebras of inequivalent observables are
  isomorphic as Poisson algebras).
\end{Cor}

We also note a difference between the initial and the extended
constraint surfaces~$\Sigma$ and~$\ZQ$ in that~$\Sigma$ carries an
action of the gauge generators~$\pb{T_i}{\cdot\,}$, while~$\ZQ$ is
equipped with the zero locus antibracket.  This is not unnatural,
because the on-shell gauge symmetries are Hamiltonian vector fields
with respect to the zero-locus antibracket, while inequivalent
observables are (identified with) the characteristic functions of the
zero-locus antibracket.\footnote{This applies at the classical level.
  The notion of the initial and the extended constraint surfaces is
  essentially classical and has no obvious counterparts at the quantum
  level.  At the quantum level, restrictions to the constraint surface
  should be understood as restriction to some quotient of the full
  Hilbert space of the quantum system.  We do not discuss this very
  interesting subject here, and refer instead to~\cite{[BM-QA]}, where
  a related problem was considered.  We thank I.A.~Batalin for an
  illuminating discussion of this point.}

Finally, we note that there is a slightly different point of view on
the interpretation of BFV observables in terms of the geometry of
$\ZQ$.  Namely, to each (odd) Poisson structure, one can associate the
coboundary operator (differential) acting on antisymmetric tensor
fields, with the action being the adjoint action of the Poisson
bivector with respect to the Schouten-Nijenhuis bracket.  Inequivalent
observables are then the \textit{zero-degree cohomology} of this
differential on $\ZQ$ (tensors of zero degree are functions).

\subsection{Observables, gauge symmetries, and zero locus reduction
  in the BV quantization}\label{sec:gaugesymm} The above can be
reformulated for odd \QP~manifolds/BV quantization.  In the BV
formulation, the zero locus of~$\Q=\ab{S}{\cdot}$, where~$S$ is the
master action, is the stationary surface of~$S$ (provided the BV
antibracket~$\ab{\,}{\,}$ is nondegenerate).  The BV observables
are the cohomology of~$\Q$ in the ghost number zero.  The BV gauge
symmetries are the vector fields of the form
\begin{equation}\label{eq:BV-gs}
  X=\ab{\Q B}{\cdot\,},
\end{equation}
and, thus, are Hamiltonian vector fields generated by trivial
observables. Whenever the master action~$S$ is constructed via the BV
prescription starting from a given initial action~$S_0$, the zero
locus of~$\Q=\ab{S}{\cdot}$ is a certain extension of the
stationary surface of the initial action~$S_0$.

At the formal level, all the statements considered in the BFV scheme
have their counterparts in the BV formalism.  We do not restate here
the contents of~\bref{sec:basic}--\bref{sec:ZQ} for the odd case and
refer instead to~\cite{[GST]}.\footnote{To avoid misunderstanding, we
  note that we have changed our point of view on how the BFV/BV gauge
  symmetries should be \textit{defined}: the BV gauge symmetries were
  called the ``trivial gauge symmetries of the master action''
  in~\cite{[GST]}; translating the results proved in~\cite{[GST]} into
  the present conventions, therefore, requires some care with the
  ``obsolete'' definitions.}  We only point out one important
difference.  Unlike the Hamiltonian picture, the Lagrangian one can be
considered in the scope of a finite dimensional analogue only
formally.  The finite dimensional configuration space (the space of
field histories) does not correspond to any physically relevant
system. Thus all the BV counterparts of the statements of the previous
section should be considered with some care.  In particular, the BV
quantization prescription requires the master action~$S$ to be a
proper solution to the master equation.  The condition imposed on the
master action to be proper has no counterpart in the Hamiltonian
picture. It implies that the corresponding configuration space is a
proper \QP~manifold (which in general is not the case for the BFV
phase space).  In the finite dimensional case, this in turn implies
that all the observables (the cohomology of~$\Q$) are trivial (except
those of a topological nature).  The~$\Q$ cohomology becomes
nontrivial only when evaluated on space-time local
functionals~\cite{[Bering],[BT-finit]}.

\section{Towers of brackets}\label{sec:towers}
In this section, we study the possibility of a ``second'' zero-locus
reduction, i.e., the reduction on a \QP~manifold which itself is the
result of a zero-locus reduction.  This leads to several well-known
structures, including the classical Yang--Baxter equation.

\subsection{A ``second'' zero-locus reduction} On a \QP~manifold
$\manN$ (which can be either even or odd), a coisotropic submanifold
$\manL \subset \ZQ$ (for example, a Lagrangian submanifold in~$\manN$)
is a P-manifold, i.e., is equipped with an (even or odd) Poisson
structure (see~\bref{thm:general-zero}).  One can try to equip~$\manL$
with a compatible~$\Q$ structure, thereby making it into a
\QP~manifold.  On a general \QP~manifold~$\manN$, there is no
canonical structure inducing a~$\Q$ operator on~$\manL$\@.  Instead,
we can look for a~$\Q$ operator on~$\manL$ in the form
$\Q_\manL=\gerst{H}{\cdot\,}_{\Q}$, where~${\gerst{~}{\;}}_{\Q}$ is
the bracket given by~\eqref{eq:LPB} and~$H$ is a solution of the
equation
\begin{equation}
  {\gerst{H}{H}}_{\Q}=0,\qquad H \in \func\manL,\quad
  \p{H}=\p{{\gerst{~}{\;}}_{\Q}}+1.
\end{equation}
Whenever such an~$H$ is found,~$\manL$ becomes a \QP~manifold.  With
this~$\Q$-structure, we can repeat the procedure, thereby producing a
sequence of \QP~manifolds.

This construction can be restated in terms of differential Poisson
algebras (the algebras of functions on \QP~manifolds).  Even ``more
algebraically,'' we consider the case where a differential Poisson
algebra arises from a \textit{complex} endowed with a
super-commutative associative multiplication and a Gerstenhaber-like
multiplication (see the Appendix).  To these differential Poisson
algebras, we can then apply one or more zero-locus reduction steps,
resulting in relations between different complexes.

\subsection{Examples of the zero locus reduction on an even
  \QP~manifold} Let~$\manM$ be a cotangent bundle~$\manM=T^*\manX$. We
then write~$(q^a,\,p_a)$ for local coordinates on~$\manM$ (which we
take to be bosonic to avoid extra sign factors); the Poisson bracket
then is $\pb{q^a}{p_b}=\delta^a_b$.  We assume a Hamiltonian action of
a Lie algebra~$\aA$ on~$\manM$.  For simplicity, we consider the
Hamiltonian action that is the lift of an action on $\manX$ via the
vector fields~$X_i=X^a_i\frac{\d}{\d q^a}$, with
$[X_i,X_j]=C^k_{ij}X_k$.  The generators of the Hamiltonian action on
~$\manM$ are then given by~$T_i=-p_a X_i^a(q)$.  Applying the BFV
scheme to the constraints~$T_i$ gives the BRST generator
\begin{equation}\label{eq:omega}
  \Omega=-p_a X^a_i(q) \theta^i -
  \thalf C^k_{ij}\xi_k\theta^i\theta^j.
\end{equation}

We now take the submanifold~$\manL\subset\ZQ$ (which is Lagrangian
in~$\manM$) determined by~$\theta^i=0$ and~$p_a=0$ and view~$q^a$ and
$\xi_i$ as local coordinates on $\manL$.  The antibracket
$(~,~)\equiv\gerst{\;}{\;}_{\Q}$ from~\bref{thm:general-zero} is then
given by
\begin{equation}\label{eq:BFV-antibracket}
  \ab{\xi_i}{\xi_j}=C^k_{ij}\xi_k,\qquad
  \ab{q^a}{\xi_i}=-X_i^a.
\end{equation} 
Using this antibracket structure on $\manL$, we consider the equation
\begin{equation}\label{eq:CYBE} 
  \ab{H}{H}=0
\end{equation}
for an even function $H\in\func\manL$.  Given a solution $H$, we can
construct the odd nilpotent vector field $\Q=\ab{H}{\cdot\,}$ that
makes $\manL$ into a \QP~manifold.

We consider solutions to~\eqref{eq:CYBE} of the form
\begin{equation}\label{YB-Hamiltonian}
  H_{\YB}=-\half r^{ij}\xi_i\xi_j,
\end{equation}
where $r$ is a skew-symmetric matrix with entries from~$\func{\manX}$.
Explicitly, Eq.~\eqref{eq:CYBE} is the following generalization of the
CYBE:
\begin{equation}\label{eq:CYBE-expl}
  r^{\ell[i}C^k_{\ell m} r^{j]m} +
  X^a_{\ell} r^{\ell[i}\frac{\d}{\d q^a} r^{jk]}=0.
\end{equation}

We now proceed with the next stage of the zero locus reduction.  The
zero locus of the ``Yang--Baxter differential''
$\Q_{\YB}=(H_{\YB},{}\cdot{})$ is determined by $r^{ij}\xi_j=0$. We
choose a smaller submanifold $\manX\subset\Z{\Q_{\YB}}$ determined by
$\xi_i=0$.  Whenever~\eqref{eq:CYBE} is satisfied,
$\pb\cdot\cdot=\ab{\cdot}{\Q_\YB\,\cdot}$ is a Poisson bracket
on~$\manX$.  Explicitly, the Poisson brackets are given by
\begin{equation}\label{eq:YB-Poisson-bracket}
  \pb{q^a}{q^b}=X^a_i r^{ij}X^b_j.
\end{equation}

\subsubsection{The classical Yang--Baxter equation}\label{sec:YB}
Antibracket~\eqref{eq:BFV-antibracket} considered on $q$-independent
functions coincides with the Schouten bracket on~$\wwedge\aA$ viewed
as the Grassmann algebra generated by~$\xi_i$.  In the case where
$r^{ij}$ is a constant matrix, \eqref{eq:CYBE-expl} becomes the CYBE
\begin{equation}\label{CYBE}
  r^{j[i}C^k_{jl}r^{m]l}=0.
\end{equation}
For each $r^{ij}$ satisfying~\eqref{CYBE}, the corresponding
differential $\Q_{\YB}$ (considered on~$\wwedge\aA$) is nothing but
the cohomology differential of the Lie algebra complex with trivial
coefficients (see Appendix~\ref{standard-definition}), for the Lie
algebra defined on~$\aA^*$ by the structure constants
$F^{ij}_k=r^{il}C^j_{lk}-r^{jl}C^i_{lk}$.

\subsubsection{The Sklyanin bracket} With $\manX$ taken to be the Lie
group corresponding to the Lie algebra $\aA$, we have two natural ways
to define the action of $\aA$ on $\manX$, by the left- and
right-invariant vector fields $L_i$ and $R_i$.  Proceeding along the
steps described in the previous paragraphs with~$X_i^a$ taken to
be~$L_i^a$ or~$R_i^a$, we arrive at two Poisson brackets on~$\manX$,
\begin{equation}
  {\pb{q^a}{q^b}}_{\text{right}}=L^a_i r^{ij}L^b_j
  \qquad\text{and}\qquad
  {\pb{q^a}{q^b}}_{\text{left}}=R^a_i r^{ij}R^b_j,
\end{equation}
\textit{which are compatible} in view of~$\commut{R_i}{L_j}=0$.  The
Poisson bracket
\begin{equation}
  {\pb{q^a}{q^b}}_{\text{Sklyanin}}=
  {\pb{q^a}{q^b}}_{\text{right}}-
  {\pb{q^a}{q^b}}_{\text{left}}
\end{equation}
makes the Lie group~$\manX$ into a Poisson--Lie group.

\subsection{Zero locus reduction on an odd QP
  manifold}\label{subsec:BV} To reformulate the above for an odd
\QP~manifold, we construct the BV scheme starting with a manifold
$\manX$ with an $\aA$ action.  The~$\xi_i$ variables are then even,
and because of the symmetry properties, the ``tower of reductions'' is
shorter than for odd~$\xi_i$.  We then introduce antifields~$q^*_a$,
ghosts~$\theta^i$, and their antifields~$\xi_i$,\pagebreak[3] with
$\ab{\theta^i}{\xi_j}=\delta^i_j$ (where restored the traditional
notation for the antibracket).  The differential
\begin{equation}\label{eq:TBV}
  \Q=(S,\cdot),\qquad
  S = q^*_a X^a_i\theta^i
  -\thalf \xi_k C^k_{ij}\theta^i \theta^j
\end{equation}
corresponds to the quantization of a theory with the vanishing
classical action.

We choose a Lagrangian subspace $\manL \subset \Z\Q$ determined by
$\theta^i=0$ and $q^*_a=0$.  In accordance with
Sec.~\bref{zero-locus}, the zero locus reduction induces a Poisson
bracket~$\pb{~}{\;}_{\Q}$ on~$\manL$ with the nonvanishing
components
\begin{equation}\label{eq:BV-PB}
  {\pb{\xi_i}{\xi_j}}_{\Q}=C^k_{ij}\xi_k,\qquad
  {\pb{q^a}{\xi_i}}_{\Q}=X_i^a.
\end{equation}
Unless $\manX$ is a \textit{super}manifold, $\manL$ is a purely even
manifold, and therefore, the new generating equation with respect to
the ${\pb{\,}{}}_{\Q}$-bracket has only the trivial solution.  The
tower of brackets is thus terminated.

We now recall that the \textit{even} variables $\xi_i$ generate the
algebra of functions on~$\aA^*$. Restricting ourselves to functions
that are independent of the coordinates on $\manX$, we see
that~\eqref{eq:BV-PB} becomes the Berezin--Kirillov bracket
on $\aA^*$,
\begin{equation}\label{bracket-linear}
  \{f,\,g\}=f\,\ddr{}{\xi_i}\,\xi_k\,C_{ij}^k\,\ddl{}{\xi_j}g.
\end{equation}

\subsubsection{Linear and nonlinear brackets} The bracket
in~\eqref{bracket-linear} is ``linear'' in the sense of its explicit
dependence on~$\xi_i$.  For a Lie algebra~$\aA$, one can construct
``nonlinear'' brackets $\ddr{}{\xi_i}\Omega_{ij}\ddl{}{\xi_j}$
on~$\aA^*$, where the expansion of $\Omega_{ij}$ in $\xi_i$ starts
with $\xi_k\,C_{ij}^k$.  For a given bracket of this form, a natural
problem is whether it can be transformed into the Berezin--Kirillov
bracket by a change of coordinates.  With the help of the zero-locus
reduction, this is solved as follows.  The Poisson bracket is
represented as the zero-locus reduction of the \textit{canonical}
antibracket on a \QP~manifold with $\Q$ determined by the Hamiltonian
$H=\Omega_{ij}(\xi)\theta^i\theta^j$.  The Jacobi identity for the
Poisson bracket is rewritten as the master equation for $H$, and
moreover, the terms containing higher powers of~$\xi_i$ are closed
with respect to the differential $\Q_0=\{H_0,{}\cdot{}\}$, where
$H_0=\xi_k\,C_{ij}^k\,\theta^i\theta^j$ is the ``linear'' part of the
Hamiltonian.  We thus have proved the fact known from other
considerations (and in a more powerful analytic
version)~\cite{[Arnold]}
\begin{Cor}
  Let
  $\Omega_{ij}(\xi)=\xi_k\,C_{ij}^k+\xi_k\xi_l\,C_{ij}^{kl}+\ldots$ be
  the matrix of a Poisson bracket on~$\aA^*$, where $C_{ij}^k$ are the
  structure constants of a Lie algebra~$\aA$.  Then~$\Omega_{ij}(\xi)$
  can be reduced to the form~$\xi_k\,C_{ij}^k$ by a change of
  variables~$\xi_i\mapsto f_i(\xi)$ if the second cohomology
  group~$H^2(\aA,\S\aA)$ of~$\aA$ with coefficients in~$\S\aA$ is
  trivial.
\end{Cor}
Similar considerations in the BFV case lead to similar statements for
the nonlinear antibracket.

\section{Bi-QP~manifolds} \label{bicomplex} Up to now, we have studied
\QP~manifolds whose differential corresponds to a \textit{single}
solution of the corresponding ``master'' equation.  We now consider
bi-\QP~manifolds.

\subsection{A BFV-like formulation of the bialgebra complex} In the
previous section, we associated an even \QP~manifold with a vector
space $\aA$ and a smooth manifold $\manM=T^*\manX$. Namely, a Lie
algebra structure on $\aA$ and the vector fields $X_i$ (giving an
$\aA$-module structure on~$\func\manX$) can be read off from a
solution of the generating equation
\begin{equation}\label{eq:BFV-master-2}
  \{\Omega,\Omega\}=0
\end{equation}
with the ansatz~\eqref{eq:omega}.  The algebra of functions on the
thus constructed \QP~manifold is $\cA=\Hom{\wwedge \aA}{\wwedge
  \aA}\tensor\func{T^*\manX}$; we interpret $\Hom{\wwedge\aA}{\wwedge
  \aA}$ as the algebra generated by the odd variables $\theta^i$ and
$\xi_j\,$.  The basic Poisson bracket relations are
\begin{equation}\label{bicomplex-Poisson}
  \pb{\theta^i}{\xi_j}=\delta^i_j,\qquad
  \pb{q^a}{p_b}=\delta^a_b,
\end{equation}
where $q,p$ are the standard local coordinates on the cotangent
bundle~$\manM=T^*\manX$.  We have the solution
\begin{equation}\label{eq:C}
  C=-p_a X_i^a(q) \theta^i - \thalf \xi_k C^k_{ij}\theta^i\theta^j.
\end{equation}
At the same time, every solution of~\eqref{eq:BFV-master-2} of the
form
\begin{equation}\label{eq:F}
  F=-p_a X^{i\,a} \xi_i - \thalf \theta^k F_k^{ij}\xi_i\xi_j
\end{equation}
determines a coalgebra structure on the vector space $\aA$, or
equivalently, a Lie algebra structure on~$\aA^*$, and makes
$\func\manX$ into an $\aA^*$-module, with the vector fields
$X^i=R^{i\,a}\diff{q^a}\in \Vect\manX$ representing the action of the
basis elements~of~$\aA^*$.  Then $\cA$ is equipped with Poisson
bracket~\eqref{bicomplex-Poisson} and the differentials
\begin{equation}\label{two-BFV-differential}
  \begin{split}
    d_C={}&\pb{C}{\cdot\ }\\
    ={}&-\thalf C^k_{ij}\theta^i\theta^j\diff{\theta^k}
    -\xi_k C^k_{ij}\theta^i\diff{\xi_j} -p_a X_i^a \diff{\xi_i}
    +\theta^i X_i^a \diff{q^a}
    -\theta^i p_a X_{i,b}^a \diff{p_b},\\
    d_F={}&\pb{F}{\cdot\ }\\
    ={}&-\thalf F_k^{ij}\xi_i\xi_j\diff{\xi_k}
    -\theta^k F_k^{ij}\xi_i \diff{\theta^j} -p_a X^{i\,a}
    \diff{\theta^i} +\xi_i X^{i\,a} \diff{q^a} -\xi_i p_a
    X_{,b}^{i\,a} \diff{p_b}.
  \end{split}
\end{equation}
We next impose the condition that the differentials be compatible,
i.e.,
\begin{equation}\label{cocycle}
  \commut{d_C}{d_F}=0\ \Longleftrightarrow\ \pb{C}{F}=0.
\end{equation}
\begin{prop}
  Condition \eqref{cocycle} implies that $(\aA,\aA^*,\aA \oplus
  \aA^*)$ is a Manin triple~\cite{[Manin]}, with the Lie bracket on
  $\aA\oplus \aA^*$ given by
  \begin{equation}
    \label{eq:triple}
    \commut{e_i}{e_j}=C^k_{ij} e_k, \qquad
    \commut{e^i}{e^j}=F_k^{ij} e^k, \qquad
    \commut{e_i}{e^j}=C^j_{ik} e^k+F^{jk}_ie_k.
  \end{equation}
  where $e_i$ and $e^i$ are dual bases in $\aA$ and $\aA^*$
  respectively.  Equivalently, $\aA$ is a Lie bialgebra.  Moreover,
  $\func\manX$ is a module over the Lie algebra~$\aA\oplus\aA^*$.
\end{prop}
\noindent
The proof is straightforward.

That $\func\manX$ is a module over $\aA\oplus\aA^*$ means that under
the mapping~$e_i\mapsto X_i$, $e^i\mapsto X^i$, the following
commutation relations between vector fields are satisfied:
\begin{equation}\label{eq:d-module}
  \commut{X_i}{X_j}=C^k_{ij}X_k,\qquad
  \commut{X^i}{X^j}=F_k^{ij}X^k, \qquad
  \commut{X_i}{X^j}=C^j_{ik}X^k+F_i^{jk}X_k.
\end{equation}
It also follows from \eqref{cocycle} that
\begin{equation}\label{eq:antisymmetry}
  X^{ia}X^b_i+X^{ib}X^a_i=0.
\end{equation}

\subsubsection{Zero locus reduction on a bi-QP~manifold} We next
consider the submanifolds of the zero loci, $\manL_{C}\subset\Z{C}$
and $\manL_{F}\subset\Z{F}$ defined by $(\theta^i=0,~p_a=0)$ and
$(\xi_i=0,~p_a=0)$, respectively.  Since $\manL_{C}$ and $\manL_{F}$
are coisotropic, we can apply Theorem~\ref{thm:general-zero}.  We thus
have the respective antibrackets
\begin{equation}\label{two-BFV-antibracket}
  \begin{split}
    \pb{\xi_i}{\xi_j}_C=&C^k_{ij}\xi_k,\qquad
    \pb{\xi_i}{q^a}_C=X_i^a,\\
    \pb{\theta^i}{\theta^j}_F=&F_k^{ij}\theta^k,\qquad
    \pb{\theta^i}{q^a}_F=X^{i\,a},
  \end{split}
\end{equation}
on $\manL_{C}$ and $\manL_{F}$.
\begin{prop}
  The differential $d_C$ induces a well-defined operator (vector
  field) $\bar d_C=d_C|_{\manL_F}:\func{\manL_{F}}\to
  \func{\manL_{F}}$ and the differential $d_F$ induces an operator
  $\bar d_F=d_F|_{\manL_C}:\func{\manL_{C}}\to \func{\manL_{C}}$.
  Thus, $\func{\manL_{F}}$ \textup{(}$\func{\manL_{C}}$\textup{)} is
  an odd differential Poisson algebra and $\manL_{F}$ (respectively,
  $\manL_{C}$\textup{)} is an odd \QP~manifold.
\end{prop}

Thus, the manifolds $\manL_C$ and $\manL_F$ are equipped with $\Q$
structures.  We now proceed to the next step of the zero locus
reduction.

Recall that the submanifold $\manX=\manL_{C}\cap \manL_{F}$ is
determined by the equations $p_a=\xi_i=\theta^j=0$.  It is easy to see
that $\manX$ is a coisotropic submanifold of $\manL_{C}$ and also a
coisotropic submanifold of~$\manL_{F}$.  On~$\manX$, we then have the
Poisson bracket
\begin{equation}
  {\pb\cdot\cdot}_{\manX}=\pb{\cdot}{\bar d_F\,\cdot}_C=
  \pb{\cdot}{\bar d_C\,\cdot}_F
\end{equation}
or in the coordinate form,
\begin{equation}\label{YB-Poisson-bracket2}
  \pb{q^a}{q^b}_{\manX}=X^a_i X^{i\,b}.
\end{equation}
It follows from \eqref{eq:antisymmetry} that bracket
\eqref{YB-Poisson-bracket2} is skew-symmetric; the Jacobi identity
follows from the compatibility of~$d_C$ and~$d_F$.

\subsubsection{Coboundary bialgebras} Up to this point, the situation
was symmetric with respect to~$\theta^i\leftrightarrow\xi_i$, but now
we try to solve Eq.~\eqref{cocycle} for~$F$.  Namely, suppose that~$F$
is a coboundary
\begin{equation}\label{coboundary}
  F=d_C r=\pb{C}{r},
\end{equation}
where $r=r^{ij}\xi_i\xi_j$ and $r^{ij}$ is taken to be a constant
matrix.  Then the condition~$d_F^2=0$ yields
\begin{equation}\label{generalized-CYBE}
  \pb{C}{\pb{r}{\pb{C}{r}}}=d_C\pb{r}{d_C\,r}=0.
\end{equation}
This is the generalized CYBE\@.  An even stronger condition
\begin{equation}
  \pb{r}{d_C\,r}=\pb{r}{r}_C=0
\end{equation}
leads to the~CYBE (see~\eqref{eq:CYBE}).

\subsection{Two differentials from a Lie algebra action} We now
look at the bicomplex setting from a somewhat different point of view.
Rather than associating a second differential with a coalgebra
structure, we construct a pair of differentials for a single Lie
algebra.  This subject attracts one's attention because of its
possibly deep relation to the extended BRST
symmetry~\cite{[BLT],[BLT-sp2]}.  We now show that the
bicomplex generalization of the zero locus reduction method induces
the non-Abelian triplectic antibrackets on the space of common zeroes
of the differentials.\footnote{The non-Abelian triplectic antibrackets
were introduced in \cite{[G]}, see also \cite{[GS]}, as the
structure underlying a possible generalization of the well known
Lagrangian version of the extended BRST quantization.}

\subsubsection{Left and right $\aA$ actions}
We consider the left and the right actions of $\aA$ on $\manX$.  To
illustrate the idea, we restrict ourselves to the case where
$\manX=\manG$ is the Lie group corresponding to the Lie algebra $\aA$.
Let the basis elements $e_i$ of $\aA$ act on $\manG$ via the left
invariant vector fields $L_i$ (which correspond to the right action)
and via the right invariant vector fields $R_i$ (which correspond to
the left action).  Obviously, $\commut{L_i}{R_j}=0$.  Let $q^a$ and
$p_a$ be the standard coordinates on~$T^*\manG$.  Unlike in the case
considered above, we introduce the doubled set of variables $\xi^1_i$,
$\xi^2_j$, $\theta_1^k$, and $\theta_2^l$, $i,j,k,l=1,\dots,\dim\aA$,
with the basic Poisson brackets
\begin{equation}
  \pb{q^a}{p_b}=\delta^a_b,\quad
  \pb{\theta_1^i}{\xi^1_j}=\delta^i_j, \quad
  \pb{\theta_2^i}{\xi^2_j}=\delta^i_j.
\end{equation}

The functions
\begin{equation}\label{eq:biOMEGA}
  \begin{split}
    \Omega^1&=-p_a R^a_i\theta_1^i
    -\thalf \xi^1_k C^k_{ij}\theta_1^i \theta_1^j,\\
    \Omega^2&=-p_a L^a_i\theta_2^i
    -\thalf \xi^2_k C^k_{ij}\theta_2^i \theta_2^j
  \end{split}
\end{equation}
satisfy $\pb{\Omega^\alpha}{\Omega^\beta}=0$ for
$\alpha,\beta=1,2$, as follows immediately from the commutativity of
the left- and right-invariant vector fields.  These generating
functions give rise to the anticommuting differentials
$\Q^a=\pb{\Omega^a}{\cdot\,}$, thereby providing
$\BFV_{\mathrm{ext}}$ with a bicomplex structure.

\subsubsection{Zero locus reduction in $\BFV_{\mathrm{ext}}$ and
  nonabelian triplectic antibrackets} We now apply the zero locus
reduction along the lines of Sec.~\bref{zero-locus}.  We identify the
zero locus $\Z{\Q^1}$ (respectively, $\Z{\Q^2}$) of the differential
$\Q^1$ (of~$\Q^2$) determined by the equations $\theta_1^i=0$ and
$p_a=0$ (respectively, $\theta_2^i=0$ and $p_a=0$).  The intersection
$\manL=\Z{\Q^1}\cap\Z{\Q^2}$ is then endowed with a pair of compatible
antibrackets.  Identifying $\func{\manL}$ (functions on the
intersection) with functions of $q^a$, $\xi^1_i$, and $\xi^2_j$, we
have
\begin{equation}\label{eq:triAB}
  \begin{split}
    \pb{\xi^1_i}{q^a}_{\Q^1}=R^a_i,\qquad
    \pb{\xi^1_i}{\xi^1_j}_{\Q^1}=C^k_{ij}\xi^1_k,\\
    \pb{\xi^2_i}{q^a}_{\Q^2}=L^a_i,\qquad
    \pb{\xi^2_i}{\xi^2_j}_{\Q^2}=C^k_{ij}\xi^2_k,
  \end{split}
\end{equation}
with all the other brackets vanishing. These are precisely the
non-Abelian triplectic antibrackets from~\cite{[G]}.

\section{Conclusions}
Our results give a geometric interpretation to a number of structures
involved in the BFV/BV formalism; the interpretation of the BRST
cohomology in terms of the constraint surface geometry~\cite{[HTS]}
can thus be extended in terms of geometry of a ``more invariant''
object---the zero locus~$\ZQ$ that plays the role of the
\textit{extended constraint surface}.  Although this is presently
limited to the ghost number zero, it would be interesting to extend
this interpretation to other ghost numbers. Another interesting
application of the zero locus reduction consists in interpreting $\ZQ$
with the induced Poisson bracket in the BV formulation of a pure-gauge
model as an extended phase space and the extended Poisson bracket in
the BFV formulation of the same model~\cite{[talk]}.  As noted above,
the zero locus reduction applies to finite-dimensional models; it
would be interesting to extend it to local field theory, for example
in the jet language formulation of the BRST formalism~\cite{[H-jet]}.

\subsubsection*{Acknowledgments} We are grateful to I.~A.~Batalin,
P.~H.~Damgaard, O.~M.~Khudaverdyan, and I.~V.~Tyutin for illuminating
discussions.  This work was supported in part by the Russian
Federation President Grant~99-15-96037.  The work of AMS and MAG was
supported in part by the RFBR Grant 99-01-00980, and the work of MAG
was also supported by the INTAS YSF-98-156. MAG is grateful to
P.~H.~Damgaard for kind hospitality at the Niels Bohr Institute, where
a part of this paper was written.

\appendix
\section{Lie algebra cohomology and the (anti)bracket}
\label{standard-definition}
Let~$\aA$ denote a Lie algebra of dimension~$N$ and~$\mM$ denote an
$\aA$-module.  We denote by
\begin{equation}\label{w-decomp}
  \wwedge\aA=\bigoplus_{n=0}^N\wedge^n\aA
\end{equation}
the exterior algebra of the vector space $\aA$ and by $\S\aA$ the
symmetric tensor algebra.

The cohomology complex of~$\aA$ with coefficients in the module~$\mM$
is
\begin{equation}\label{standard-cohomology}
  \C^*(\aA,\mM)=\{\Hom{\wwedge\aA}{\mM},~d\}.
\end{equation}
Decomposition~\eqref{w-decomp} induces the grading \ 
$\C^*(\aA,\mM)=\bigoplus_{n=0}^N C^n(\aA,\mM)$, \ where \ 
$C^n(\aA,\mM)=\\\Hom{\wedge^n\aA}{\mM}$. The differential~$d$ has the
degree~$1$ and acts as $d\,:\,C^n(\aA,\mM)\rightarrow
C^{n+1}(\aA,\mM)$ via
\begin{multline}\label{coh-diff}
  (da)(g_1,\ldots,g_{n+1})= \sum\limits_{1 \leq i < j \leq n+1}
  (-1)^{i+j-1} a(\Lie{g_i}{g_j},g_1,\ldots,\hat g_i,\ldots,\hat
  g_j,\ldots,g_{n+1})+\\
  +\sum\limits_{1 \leq i \leq n+1} (-1)^{i} g_i a(g_1,\ldots,\hat
  g_i,\ldots,g_{n+1}),\qquad a\in C^n(\aA,\mM).
\end{multline}
We also use the simplified notation~$C^n=C^n(\aA,\mM)$.

We can identify the cohomology complex~$\C^*(\aA,\mM)$
with~$\wwedge\aA^*\tensor\mM$ as follows.\footnote{We here assume that
  the algebras are finite dimensional or graded, $\aA=\oplus_i\aA_i$,
  with finite dimensional homogeneous spaces~$\aA_i$, and~$\aA^*$ is
  by definition~$\aA^*=\oplus_i\aA^*_i$, where~$\aA^*_i$ are finite
  dimensional spaces dual to~$\aA_i$.}  Let $e_i$ be a basis in~$\aA$,
with~$\Lie{e_i}{e_j}=C_{ij}^k e_k$.  Let also~$\theta^i$ be the basis
of~$\aA^*$ dual to~$e_i$.  The Grassmann algebra generated by
$\theta^i$ is then identified with $\wwedge\aA^*$.  To every cochain
$x\in\Hom{\wedge^n\aA}{\mM}$, we associate the element (with the
summations implied)
\begin{equation}\label{bar-a}
  \bar{x}=\frac{1}{n!}\,x(e_{i_1},\dots,e_{i_n})\,
  \theta^{i_1}\dots\theta^{i_n}
  \in\wwedge\aA^*\tensor\mM.
\end{equation}
The differential~$d$ then acts on $\wwedge\aA^* \tensor \mM$ as the
differential operator
\begin{equation}\label{eq:d-1}
  d=\thalf C^k_{ij}\theta^i\theta^j\diff{\theta^k}-\theta^i X_i,
\end{equation}
where $X_i\,:\,\mM\to\mM$ is the action of $e_i \in \aA$ on $\mM$.

We next specialize to the coefficients in~$\aA$ (viewed as the adjoint
representation $\aA$-module).  The complex is then endowed with the
Gerstenhaber bracket~\cite{[Gerst1]},\cite[and references
therein]{[Gerst2]}
\begin{equation*}
  \gerst\cdot\cdot\,:\,C^n\tensor C^m\rightarrow C^{n+m-1}
\end{equation*}
given by
\begin{equation}\label{gerst}
  \gerst{x}{y}=x \circ y-(-1)^{(m+1)(n+1)}y\circ x,\qquad
  x\in C^n,\quad y\in C^m
\end{equation}
where
\begin{equation}\label{dot-multiply}
  (x\circ y)(a_1,\dots,a_{n+m-1})=
  \tfrac{1}{m!(n-1)!}\sum_{\sigma\in P_{n+m-1}}
  (-1)^\sigma x(a_{\sigma(1)},\dots,a_{\sigma(n-1)}, y(a_{\sigma(n)},
  \dots,a_{\sigma(n+m-1)}))
\end{equation}
This makes $\Hom{\wwedge \aA}{ \aA}$ into a graded differential Lie
algebra.

Let $\xi_i$ denote the basis of~$\aA$ viewed as an $\aA$-module
(equivalently, \textit{coordinates} on $\aA^*$).  For each cochain
$x\in C^n$, we then expand $\bar{x}$ from~\eqref{bar-a} as
\begin{equation}\label{a-underline}
 \bar{x}=
  \frac{1}{n!}\,\xi_j\,x^j(e_{i_1},\dots,e_{i_n})\,
  \theta^{i_1}\dots\theta^{i_n}
\end{equation}
and rewrite the Gerstenhaber bracket as
\begin{equation}\label{eq:BRSTgerst}
  \gerst{\bar{x}}{\bar{y}}=
  \bar{x}\circ\bar{y}-(-1)^{(k+1)(l+1)}
  \bar{y} \circ\bar{x}, \qquad
  \bar{x}\circ\bar{y}=
  \bar{x}\ddr{}{\theta^i}\ddl{}{\xi_i}\bar{y},\qquad
  x\in C^k,\quad y\in C^l,
\end{equation}
where $\ddr{}{\theta^i}$ is the (right) derivative in the Grassmann
algebra and the $\ddl{}{\xi_i}$ operation is defined on the elements
of form~\eqref{a-underline} as the contraction with the element
$\xi^*_i$ of the dual basis in~$\aA^*$.  The differential then becomes
\begin{equation}\label{diff-Ham}
  d=\gerst{-\thalf C^k_{ij}\xi_k\theta^i\theta^j}{\cdot{}}=
  \thalf C^k_{ij}\theta^i\theta^j\diff{\theta^k}
  - \xi_k\theta^iC^k_{ij}\diff{\xi_j}.                     
\end{equation}
On the elements $\underline{a}$ as in~\eqref{a-underline}, the second
term represents the adjoint action (in accordance with the above
choice $\mM=\aA$).  Equation~\eqref{eq:BRSTgerst} suggests the
interpretation of a Poisson/Batalin--Vilkovisky bracket.  As it
stands, however,~\eqref{eq:BRSTgerst} can be neither of these, since
no associative supercommutative multiplication has been defined on the
cochains.\footnote{Superficially, the bracket in~\eqref{eq:BRSTgerst}
  has the grade~$-1$ since it maps as $C^m\times C^n\to C^{m+n-1}$,
  however the gradings of all the terms in the complex can be shifted
  by~$1$, after which the bracket becomes a grade-$0$ operation. On
  the other hand, an associative graded commutative multiplication
  defined on the complex would fix the grading,
  and~\eqref{eq:BRSTgerst} would become either the Batalin--Vilkovisky
  or the Poisson bracket.}  There are two remarkable possibilities to
embed $\C^*(\aA,\mM)=\C^*(\aA,\aA)$ into a complex endowed with a
multiplication: the complex
\begin{equation}
  \C^*(\aA,\S\aA)=\wwedge\aA^*\tensor\S\aA
\end{equation}
corresponding to the BV quantization, or the complex
\begin{equation}
  \C^*(\aA,\wwedge\aA)=\wwedge\aA^*\tensor\wwedge\aA
\end{equation}
corresponding to the BFV quantization.  Geometrically, these two
possibilities correspond to even and odd \QP~manifolds (see
Definition~\bref{def:QP}).

Choosing $\mM=\S\aA$, we have the complex
$\bigoplus_{m,n}\Hom{\wedge^m \aA}{{\S}^n \aA}$, which can be viewed
as the associative supercommutative algebra generated by the variables
$\theta^i$ and $\xi_j$ satisfying $\xi_i\xi_j-\xi_j\xi_i=0$,
$\theta^i\theta^j+\theta^j\theta^i=0$, and
$\theta^i\xi_j-\xi_j\theta^i=0$.\footnote{\label{foot:Grassmann}These
  relations between $\theta$ and $\xi$ variables correspond to the
  case (tacitly implied in most of our formulae) where $\aA$ is a Lie
  algebra, \textit{not} a superalgebra; then the Grassmann parities
  are simply $\p{\xi_i}=0$ and $\p{\theta^i}=1$.  However, if $\aA$ is
  a Lie superalgebra, let $\p{e_i}=\varepsilon_i$ be the Grassmann
  parities of its generators.  Then $\p{\xi_i}=\varepsilon_i$ and
  $\p{\theta^i}=\varepsilon_i+1$, and therefore,
  $\xi_i\xi_j-(-1)^{\varepsilon_i\varepsilon_j}\xi_j\xi_i=0$,
  $\theta^i\theta^j-(-1)^{(\varepsilon_i+1)(\varepsilon_j+1)}
  \theta^j\theta^i=0$, and
  $\xi_i\theta^j-(-1)^{\varepsilon_i(\varepsilon_j+1)}
  \theta^j\xi_i=0$.} It then follows that~\eqref{eq:BRSTgerst} can be
extended to an odd bracket on this complex.  The differential extends
to $\Hom{\wwedge\aA}{\S\aA}$ by the same formula
$d=\gerst{C_0}{\cdot\,}$, $C_0=-\half C^k_{ij}\xi_k\theta^i\theta^j$.
The complex is endowed with the grading known as the \textit{ghost
  number} in the BV quantization or as the Weyl complex grading in
homology theory: for a cochain $x\in\Hom{\wedge^m \aA}{\S^n \aA}$, one
has~$\gh{x}=m-2n$.
  
On the other hand, taking the coefficients to be the \textit{exterior}
algebra $\wwedge\aA$, we can extend~\eqref{eq:BRSTgerst} to an even
bracket.  With $\wwedge\aA$ identified with the algebra generated by
$\xi_i$ viewed as \textit{anticommuting variables} (with obvious
modifications in the case where $\aA$ is a Lie \textit{super}algebra,
see footnote~\ref{foot:Grassmann}), the bracket becomes the Poisson
bracket on the space $\wwedge\aA^*\tensor\wwedge\aA$ (which is
identified with functions of~$\theta^i$ and~$\xi_j$; we also assume
that $\xi_i\theta^j+\theta^j\xi_i=0$ in addition to
$\xi_i\xi_j+\xi_j\xi_i=0$).  The ghost number grading on this complex
taken from the BFV quantization is $\gh{x}=m-n$ for an element
$x\in\Hom{\wedge^m \aA}{\wedge^n \aA}$.

The coefficients can be further extended (cf.~\cite{[Gerst2]}) by
$\mM=\func\manM$, the algebra of smooth functions on a
manifold~$\manM$ such that~$\aA$ acts on $\func\manM$ \textit{by
  derivations} (vector fields on $\manM$).  We write $X_i$ for the
image of the basis elements of~$\aA$ in~$\Vect{\manM}$.  In accordance
with the BRST paradigm, one wishes the vector fields representing the
action of~$\aA$ on~$\manM$ to be Hamiltonian with respect to a bracket
structure.  For \textit{even}~$\xi_i$, this can be achieved by
replacing $\manM$ with the odd cotangent bundle~$\Pi T^*\manM$ and,
thus, the algebra~$\func\manM$ with the algebra~$\func{{\Pi
    T^*\manM}}$ of smooth functions on the odd cotangent bundle.  Then
each vector field $V=V^a \ddl{}{q^a}$ on~$\manM$ is
generated by the canonical antibracket structure on ${\Pi T^*\manM}$;
the action of the basis elements $X_i=X_i^a \ddl{}{q^a}$ on
functions is given by the antibracket
\begin{equation}\label{eq:xaction}
  X_i F=-\gerst{X^a_i q^*_a}{F}, \qquad F\in \func\manM,
\end{equation}
with $q^*_a$ being the standard coordinates on the fibers of $\Pi
T^*\manM$ (and the standard antibracket given by
$\gerst{q^a}{q^*_b}=\delta^a_b$).

For odd $\xi_i$, similarly, we can consider the functions $\func{T^*
  \manM}$ on the cotangent bundle, which allows the action of $\aA$ to
be implemented by the bracket on $\func{T^* \manM}$ (the same
formula~\eqref{eq:xaction} for the bracket, where now $q^*_a$ are
the canonical coordinates on the fibers of~$T^*\manM$).  

We note, however, that the differential 
\begin{equation}\label{eq:Cdiff}
  d=\gerst{-\thalf C^k_{ij}\xi_k\theta^i\theta^j}{\cdot\,}
  +\theta^i \gerst{X_i^a
    q^*_a}{\cdot\,}
\end{equation}
in either of the complexes
\begin{align}
  \label{eq:BV-complex}
  \C^*_{\mathrm{odd}}(\aA,\manM)={}&
  \C^*(\aA,\S\aA)\tensor\func{{\Pi T^* \manM}},\\
  \C^*_{\mathrm{even}}(\aA,\manM)={}& \C^*(\aA,\wwedge
  \aA)\tensor\func{{T^* \manM}}
  \label{eq:BFV-complex}
\end{align}
is \textit{not} compatible with the bracket.  Remarkably, the
compatibility can be achieved by changing the differentials such that
\eqref{eq:BV-complex} and \eqref{eq:BFV-complex} become the well-known
BV and BFV complexes used in the Lagrangian and Hamiltonian
quantization of gauge theories.  The term to be added to the
differential is the Koszul differential involving precisely the same
``auxiliary'' variables $\xi_i$ that were originally introduced to
rewrite the Gerstenhaber bracket in the ``geometric'' form.

To conclude, we note that we have given a homological interpretation
of the structures appearing in the BRST quantization in the example of
a Lie algebra structure (i.e., in the case where the constraints or
gauge generators form a Lie algebra).  In the most general setting,
the BRST charge and the master action in the BFV and BV cases,
respectively, can be considered as the generating functions for the
$L_{\infty}$ algebras~\cite{[Stasheff]} (see also~\cite{[AKSZ]}).
{}From this general standpoint, the Lie algebra structure appears as a
particular case.

\end{document}